\documentclass[final,3p,times,twocolumn]{elsarticle}
\usepackage{graphics,color}
\usepackage{graphicx}
\usepackage{epsfig}

\usepackage{amssymb}
\journal{Physica C}

\begin{document}

\begin{frontmatter}


%%\title{Title\tnoteref{label1}}

\title{Band filling effects on coherence lengths and penetration depth
in the two-orbital negative-$U$ Hubbard model of
superconductivity}


\author[Lublin]{Grzegorz Litak}
\author[Tartu]{Teet \"Ord}
\author[Tartu]{K\"ullike R\"ago}
\author[Tartu]{Artjom Vargunin}

\address[Lublin]{Lublin University of Technology, Faculty of Mechanical Engineering, Nadbystrzycka 36,
PL-20618 Lublin, Poland}
\address[Tartu]{Institute of Physics, University of Tartu, T\"{a}he 4, 51010 Tartu, Estonia}

\begin{abstract}
 The two-orbital superconducting state is modeled by  on-site intra-orbital negative-$U$ Hubbard
correlations together with inter-orbital pair-transfer interactions.
 The critical temperature is mainly governed by intra-orbital attractive interactions and it can pass through an additional
 maximum as a function of band filling. For the certain number of electrons the clear interband proximity
effect is observable in the superconducting state of the band with a smaller gap.
 The influence of band fillings and orbital site energies on the temperature dependencies of two-component
 superconductivity coherence lengths and magnetic field penetration depth is analyzed. The presence of proximity
effect is probably reflected in the relative temperature behaviour of characteristic lengths.
\end{abstract}


\end{frontmatter}

% \linenumbers

\section{Introduction}

 The multi-component microscopic nature of superconductivity has been established in a number of materials
(MgB$_{2}$ \cite{MgB2}, cuprates \cite{cuprates}, strontium rutenates \cite{rutenates}, iron-arsenic
 compounds \cite{pnictides}, Nb-doped SrTiO$_{3}$ \cite{str-titanate}, NbSe$_{2}$ \cite{NbSe2}, V$_{3}$Si
 \cite{V3Si} etc.). The first relevant theoretical scenarios have been discussed in the literature long time ago
 \cite{smw, m, kondo} and later on they have been developed for various systems (see e.g. Refs. \cite{kw, kko,
bianconi1, bianconi2, annett} and references therein).

 The peculiar properties of multi-component superconductors appear due to the involvement of coupled electron
 subsystems (orbitals, different Fermi surface sheets or other elements of electron structure) into the formation of
 superconducting ordering varying substantially the physical nature of the phenomenon and opening the possibilities
 for novel effects. As a result of the presence of inter-component pairing the certain quantities (e.g. coherence
 lengths and the relaxation times of superconductivity fluctuations) cannot be attributed to the initially
 independent subsystems and they describe the joint features of the whole two-component condensate \cite{babaev,
ord1, litak2012, ord2}.

 In a two-orbital superconductor with inter-orbital pairing, similarly to the results of Refs. \cite{ord1, babaev1,
 babaev2}, one of two solutions for coherence lengths diverges near the phase transition point as a function of
 temperature, while the other one is non-critical \cite{litak2012}. We note also that the existence of critical and
 non-critical length scales in the coherency associates with the conception of critical and non-critical
fluctuations \cite{ord2, ord3} in a two-component superconductor damping in different time-scales.

 In the present contribution we continue to examine the temperature behaviour of characteristic lengths of a
 two-orbital superconductor described by negative-$U$ Hubbard model \cite{Ref mrr} with inter-orbital pair-transfer
 interaction in dependence on the number of electrons. Our preliminary study has been done focusing on band filling
 effects in the spatial coherency including  Van Hove singularity \cite{litak2012}. However, in that work
 \cite{litak2012} the orbital energy differences and charge fluctuations were neglected. In the present paper we
 clarify their role in the scaling of critical and non-critical coherence lengths  and magnetic field penetration
 depth by band fillings. The numerical calculations have been carried out for a two-dimensional square lattice.

\section{Superconductivity in a two-orbital system}

Two-orbital superconductivity is modeled by the Hamiltonian of the following form (see e.g. \cite{Ref cw}):
%eq1

\begin{eqnarray}\label{1}
\!\!\!\!\!\! H&=&\sum_{\alpha}\sum_{i,j}\sum_{\sigma}\left[t_{ij}^{\alpha\alpha}+\left(\varepsilon_{\alpha}^{0}-\mu\right)
\delta_{ij}\right]a_{i\alpha\sigma}^{+}a_{j\alpha\sigma} \nonumber \\
&+&\frac{1}{2}\sum_{\alpha}\sum_{i}\sum_{\sigma}U^{\alpha\alpha} n_{i\alpha\sigma}n_{i\alpha-\sigma}\nonumber \\
&+&\frac{1}{2}\sum_{\alpha,\alpha'}^{,}\sum_{i}\sum_{\sigma}U^{\alpha\alpha'}a_{i\alpha\sigma}^{+}a_{i\alpha'\sigma}
a_{i\alpha-\sigma}^{+}a_{i\alpha'-\sigma} \, , \nonumber \\
\end{eqnarray}
where $a_{i\alpha\sigma}^{+}$ ($a_{i\alpha\sigma}$) is the electron creation (destruction) operator in the orbital $\alpha=1,2$ localized at the
site $i$; $\sigma$ is the spin index; $t_{ij}^{\alpha\alpha}$ is the hopping integral; $\varepsilon_{\alpha}^{0}$ is the orbital energy; $\mu$
is the chemical potential; $U^{\alpha\alpha}<0$ is the intra-orbital attraction energy;
$n_{i\alpha\sigma}=a_{i\alpha\sigma}^{+}a_{i\alpha\sigma}$ is the particle number operator; $U^{\alpha\alpha'}$ with $\alpha\neq\alpha'$ is the inter-orbital interaction energy.
Note that the both intra- and interorbital interaction channels involved lead to superconducting ordering.

In the reciprocal space the Hamiltonian (\ref{1}) can be represented as
%eq2
\begin{eqnarray}\label{2}
\!\!\!\!\!\!H&=&\sum_{\alpha}\sum_{\mathbf{k}}\sum_{\sigma}\left[\varepsilon_{\alpha}(\mathbf{k})-\mu
\right]a_{\alpha\mathbf{k}\sigma}^{+}a_{\alpha\mathbf{k}\sigma} \nonumber \\
&+& \frac{1}{2N}\sum_{\alpha,\alpha'}\sum_{\mathbf{k},\mathbf{k}'}\sum_{\mathbf{q}}\sum_{\sigma}U^{\alpha\alpha'}
a_{\alpha\mathbf{k}\sigma}^{+}
a_{\alpha(-\mathbf{k}+\mathbf{q})-\sigma}^{+}  \nonumber \\
&\times& a_{\alpha'(-\mathbf{k}'+\mathbf{q})-\sigma}
a_{\alpha'\mathbf{k}'\sigma} \, .
\end{eqnarray}
where $\varepsilon_{\alpha}(\mathbf{k})$ is the electron band energy associated with the orbital $\alpha$ and $N$ is the number of lattice sites (number
of atoms). The bulk superconducting state is described on the basis of the Hamiltonian (\ref{2}) by means of the Hartree-Fock-Gorkov self-consistent equations
%eq3
\begin{eqnarray} \label{3}
\!\!\!\!\!\!\Delta_{\alpha}&=&N^{-1}\sum_{\alpha'}U^{\alpha\alpha'}\sum_{\mathbf{k}}\left\langle a_{\alpha'-\mathbf{k}\downarrow} a_{\alpha' \mathbf{k}\uparrow}\right\rangle \nonumber \\
&=&\frac{-1}{N}\sum_{\alpha'}U^{\alpha\alpha'}\sum_{\mathbf{k}}\frac{\Delta_{\alpha}}{2E_{\alpha}(\mathbf{k})}
\tanh\frac{E_{\alpha}(\mathbf{k})}{2k_{B}T} \, , \nonumber \\
\end{eqnarray}
%
%eq4
\begin{eqnarray} \label{4}
\!\!\!\!\!\!n_{\alpha}&=&N^{-1}\sum_{\mathbf{k}}\sum_{\sigma}\left\langle a_{\alpha\mathbf{k}\sigma}^{+}a_{\alpha\mathbf{k}\sigma}\right\rangle \nonumber \\
&=&\frac{1}{N}\sum_{\mathbf{k}}\left[1
-\frac{\tilde{\varepsilon}_{\alpha}(\mathbf{k})}{E_{\alpha}(\mathbf{k})}\tanh\frac{E_{\alpha}(\mathbf{k})}{2k_{B}T}\right] \, , \nonumber \\
\end{eqnarray}
\begin{equation} \label{4a}
\sum_{\alpha}n_{\alpha}=n \, .
\end{equation}
Here $E_{\alpha}(\mathbf{k})=\sqrt{\tilde{\varepsilon}^{2}_{\alpha}(\mathbf{k})+\left|\Delta_{\alpha}\right|^{2}}$,
$\tilde{\varepsilon}_{\alpha}(\mathbf{k})=\varepsilon_{\alpha}(\mathbf{k})+\frac{1}{2}U^{\alpha\alpha}n_{\alpha}-\mu$,
$n_{\alpha}$ are the occupation numbers of orbitals (bands) per lattice site, and $n$ is the total number of electrons per site. The Eqs. (\ref{3}) determine the superconducting gaps of orbitals (bands), the Eq. (\ref{4}) gives us the orbital occupation numbers and the Eq. (\ref{4a}), the chemical potential $\mu$ in dependence on $n$ and other relevant quantities. The temperature of superconducting phase transition $T_{c}$ satisfies the equation
%eq5
\begin{equation}\label{5}
\left|%
\begin{array}{cc}
  1+U^{11}g_{1}\left(T_{c}\right) & U^{12}g_{2}\left(T_{c}\right) \\
  U^{21}g_{1}\left(T_{c}\right) &  1+U^{22}g_{2}\left(T_{c}\right)\\
\end{array}%
\right|\;
=0 \, ,
\end{equation}
where
%eq6
\begin{eqnarray}\label{6}
g_{\alpha}(T)= \frac{1}{2N}\sum_{\mathbf{k}}\frac{1}{\tilde{\varepsilon}_{\alpha}(\mathbf{k})}
\tanh\frac{\tilde{\varepsilon}_{\alpha}(\mathbf{k})}{2k_{B}T}  \, .
\end{eqnarray}
%Eq19

In the general case of spatial inhomogeneity and non-zero magnetic field we have the following Ginzburg-Landau equations
%eq7
\begin{eqnarray}\label{7}
\Delta_{\alpha }(\mathbf{r})&=&-\sum_{\alpha '}U^{\alpha\alpha'}\biggl[g_{\alpha'}(T)-\nu_{\alpha'}
\left|\Delta_{\alpha '}(\mathbf{r})\right|^{2} \nonumber \\
&+&\beta_{\alpha '}\left(\nabla +i\frac{2\pi}{\Phi_{0}}\mathbf{A}\right)^{2}\biggr]\Delta_{\alpha '}(\mathbf{r}) \, .
\end{eqnarray}
%
%eq8
\begin{eqnarray} \label{8}
\mathbf{j}_{s}&=&-i\frac{2\pi}{V_{0}\Phi_{0}}
\sum_{\alpha}\beta_{\alpha}\left\{\Delta^{\ast}_{\alpha}(\mathbf{r})\nabla\Delta_{\alpha}(\mathbf{r})\right.\nonumber \\
&-& \left.\Delta_{\alpha}(\mathbf{r})\nabla\Delta^{\ast}_{\alpha}(\mathbf{r})\right\}\nonumber \\
&-&\frac{2}{V_{0}}\left(\frac{2\pi}{\Phi_{0}}\right)^{2}
\left(\sum_{\alpha}\beta_{\alpha}|\Delta_{\alpha}(\mathbf{r})|^{2}\right)\mathbf{A} \, .
\end{eqnarray}
Here
%eq9
\begin{eqnarray}\label{9}
\nu_{\alpha}
 &=& \frac{-1}{2N}\sum_{\mathbf{k}}\frac{\partial}{\partial
\left|\Delta_{\alpha}\right|^{2}}\left[\frac{1}{E_{\alpha}(\mathbf{k})} \right. \nonumber \\
&\times & \left. \tanh\frac{E_{\alpha}(\mathbf{k})}{2k_{B}T_{c}}\right]_{\Delta_{\alpha}=0}  \, ,
\end{eqnarray}
%
%eq10
\begin{eqnarray}\label{10}
\beta_{\alpha}=\beta_{\alpha 1}=\beta_{\alpha 2}=\beta_{\alpha 3}
\end{eqnarray}
with
%eq11
\begin{eqnarray}\label{11}
\beta_{\alpha l}&=& \frac{-1}{4N}\sum_{\mathbf{k}}
\frac{\partial^{2}}{\partial q_{l}^{2}}\left\{
\frac{1}{\tilde{\varepsilon}_{\alpha}(\mathbf{k})+\tilde{\varepsilon}_{\alpha}(\mathbf{k}-\mathbf{q})}\right . \nonumber \\
 &\times& \left[\tanh\left(\frac{\tilde{\varepsilon}_{\alpha}(\mathbf{k})}{2k_{B}T_{c}}\right) \right. \nonumber \\
 &+& \left. \left. \tanh\left(\frac{\tilde{\varepsilon}_{\alpha}(\mathbf{k}-\mathbf{q})}{2k_{B}T_{c}}\right)\right]\right\}_{\mathbf{q}=0} \, ,
\end{eqnarray}
$\mathbf{A}$ is the vector potential, $\mathbf{j}_{s}$ is the density of supercurrent, $\Phi_{0}$ is the magnetic flux quantum and $V_{0}$ is the volume of unit cell. According to the Eq. (\ref{10}) we restrict ourselves to the isotropic electron energy spectrum.

%fig1
\begin{figure}[h]
%\begin{center}
\includegraphics[width=55mm,angle=-90]{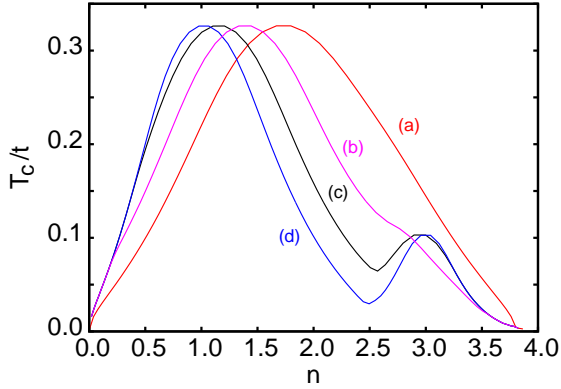}
 \caption{The critical temperature $T_c$ versus band filling $n=n_1+n_2$ for various
$ \varepsilon^{0}_1/t =
0, 1, 2, 3$, curves (a)--(d), respectively.}
%\end{center}
\end{figure}

\section{Characteristic lengths of a two-orbital superconductor}

By considering the solutions of Eqs. (\ref{7}) which correspond to the small deviations of gaps from their bulk values, $\eta_{\alpha}(\mathbf{r})=\Delta_{\alpha}(\mathbf{r})-\Delta_{\alpha}$, one can find two length scales characterizing the spatial
behaviour of superconductivity in a two-orbital system:
%eq12
\begin{eqnarray}\label{12}
\xi_{s,r}^{2}\left(T\right)=\frac{G\left(T\right)
\pm\sqrt{G^{2}\left(T\right)-4K\left(T\right)\gamma}}{2K\left(T\right)} \, ,
\end{eqnarray}
where
%eq13
\begin{eqnarray}\label{13}
G\left(T\right)&=&\left(U^{12}\right)^{2}\left[\tilde{g}_{1}(T)\beta_{2}+\tilde{g}_{2}(T)\beta_{1}\right] \nonumber\\ &-&\left[1+U^{11}\tilde{g}_{1}(T)
\right]U^{22}\beta_{2} \nonumber\\
&-&\left[1+U^{22}\tilde{g}_{2}(T)\right]U^{11}\beta_{1} \, ,
\end{eqnarray}
%
%eq14
\begin{eqnarray}\label{14}
K\left(T\right)&=&\left[1+U^{11}\tilde{g}_{1}(T)\right]\left[1+U^{22}\tilde{g}_{2}(T)\right] \nonumber\\
&-&\left(U^{12}\right)^{2}\tilde{g}_{1}(T)\tilde{g}_{2}(T) \,
\end{eqnarray}
with $\tilde{g}_{\alpha }(T)=g_{\alpha}(T)-3\nu_{\alpha}\left(\Delta_{\alpha }(T)\right)^{2}$, and
%eq16
\begin{eqnarray}\label{15}
\gamma=\left[U^{11}U^{22}-\left(U^{12}\right)^{2}\right]\beta_{1}\beta_{2} \, .
\end{eqnarray}
The soft or critical coherence length $\xi_{s}(T)$ diverges at the phase transition point $T=T_{c}$, while the rigid or non-critical coherence
length $\xi_{r}(T)$ remains finite.

On the basis of Eq. (\ref{8}) one obtains in a standard way the magnetic field penetration depth for a two-orbital superconductor,
%eq2
\begin{equation} \label{16}
\lambda=\sqrt{\frac{V_{0}\Phi^{2}_{0}}{8\pi^{2}\mu_{0}\sum\limits_{\alpha}\beta_{\alpha}|\Delta_{\alpha}|^{2}}} \, ,
\end{equation}
where $\mu_{0}$ is the magnetic permeability of free space.

\section{Numerical results}

 Below we present the results of numerical calculations for two-dimensional square lattice with hopping integrals between nearest neighbours $t^{\alpha\alpha}_{ij}=t$ and electron band
 energies associated with $s$-orbitals
 $\varepsilon_{\alpha}(\mathbf{k})=\varepsilon^0_{\alpha}-2t\left[\cos (ak_{x}) + \cos
(ak_{y})\right]$, $a$ is the lattice
 constant, and $-\frac{\pi}{a}\leq k_{x,y}\leq \frac{\pi}{a}$. The intra- and interorbital interactions are chosen
as $U^{11}=-1.5 t$, $U^{22}=-2.5 t$, $|U^{12}|=|U^{21}|=0.04 t$, and
the site energy $\varepsilon_2$ is fixed at $\varepsilon^{0}_2=0$. In all calculations we have
chosen
$k_{B}=1$.

Using Eqs. (\ref{5}) and (\ref{6}) we calculated the superconducting critical temperature $T_c$. Fig. 1 shows the dependence of $T_c$ on the band filling $n=n_1+n_2$.
One can see that the curve $T_{c}$ vs $n$ is most symmetric as
 $\varepsilon_1^0=\varepsilon_2^0$ (see Fig1, curve (a)). In this case, the small shift of the maximum of $T_{c}$ towards the smaller values of
 occupation number $n$ is caused by effective Hartree corrections $U^{\alpha \alpha} n_{\alpha}/2$. First of all, these corrections are different for various
 orbitals as the intra-orbital interactions $U^{\alpha \alpha}$ are different.  Consequently, the distances between the chemical potential $\mu$ position and the lower
edges of orbital bands are slightly different.
Therefore the position of the maximum of $T_c$ is moved reflecting the above mentioned difference which appears here due to the choice of intra-orbital pairing potentials
$|U^{22}| > |U^{11}|$.

 By increasing the difference between the bare site energies $|\varepsilon_1^0-\varepsilon_2^0|$ we observe that
 the asymmetry of the function $T_c(n)$ increases. In the cases (a) and (b) in Fig. 1, the site energies of orbitals are closer and the influence of the Hartree terms
 does not lead to dramatically different distances between the chemical potential and lower band edges. As a result the single peak in $T_c$ vs $n$ is preserved.
However, with the further increase of the difference between of
 bare site energies $\varepsilon^{0}_{1,2}$ the effect of the Hartree renormalization $|\varepsilon_1^0 + U^{11}n_1/2-\varepsilon_2^0-U^{22}n_2/2|$ becomes stronger
and we observe the splitting of the
 single peak of $T_c$ into two maxima (see Fig1, curves (c) and (d)) caused by the circumstance that the chemical potential passes in the region $0\leq n \leq 4$ two
 Van Hove singularities related to the bands $\alpha=1,2$. It should be also noted that the larger values of $|\varepsilon_1^0-\varepsilon_2^0|$ mean the larger difference
between the band occupation numbers $n_{1}$ and  $n_{2}$ which corresponds to the stronger Hartree renormalization.

%fig2
\begin{figure}[!h]
%\begin{center}

\includegraphics[width=39mm]{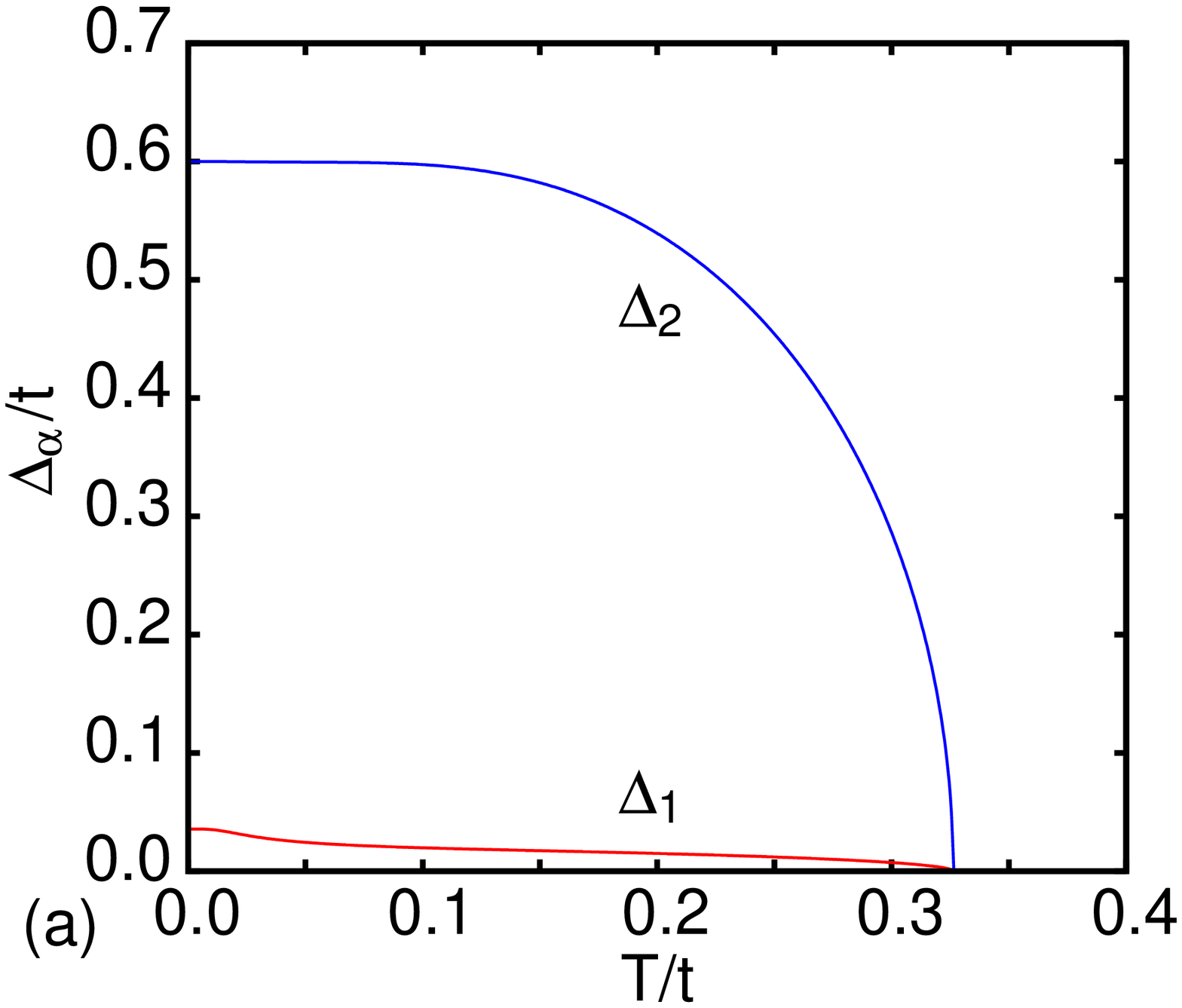} \hspace{-0.3cm}
\includegraphics[width=39mm]{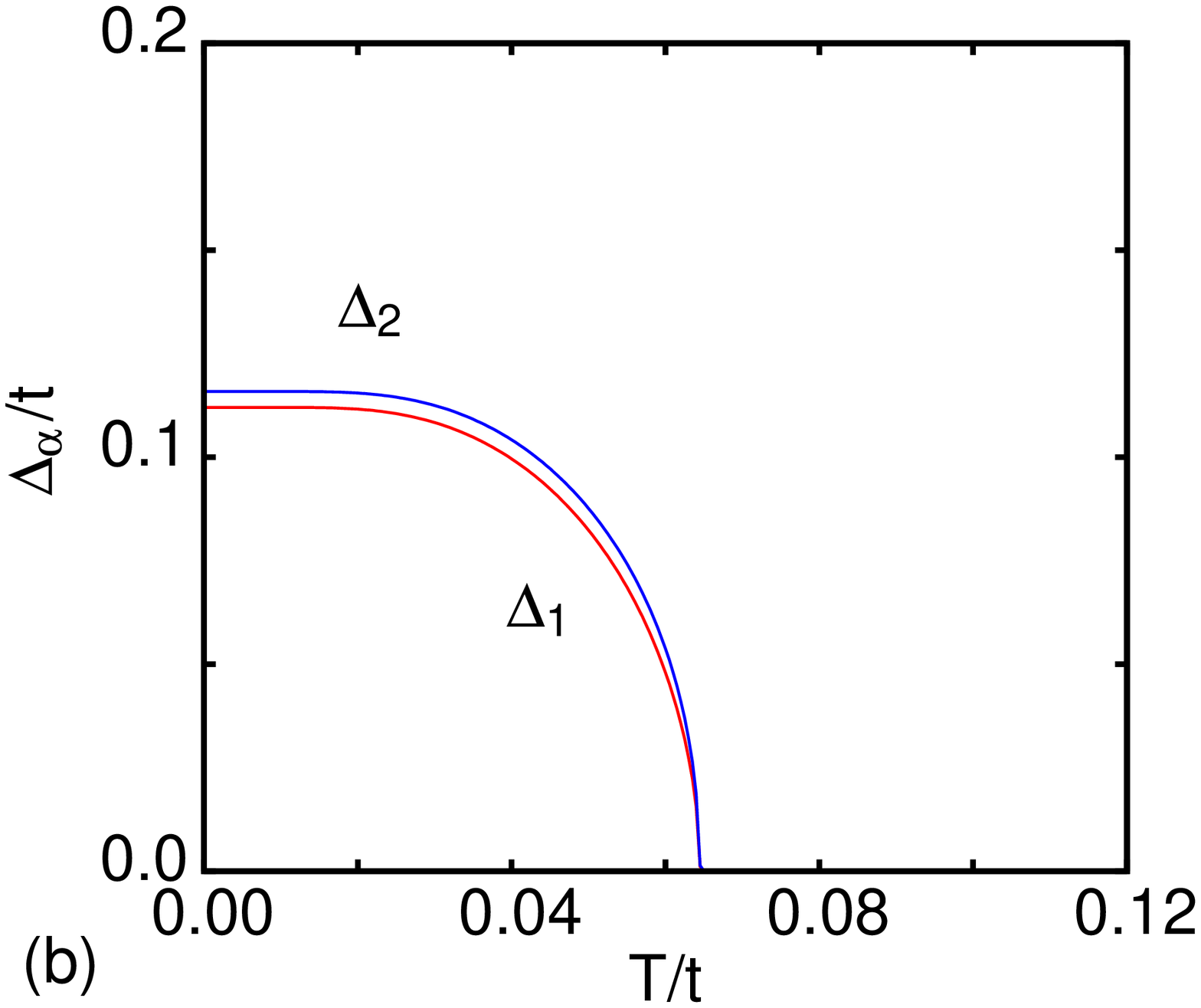}

\vspace{-1cm} \hspace{0cm}
\includegraphics[width=39mm]{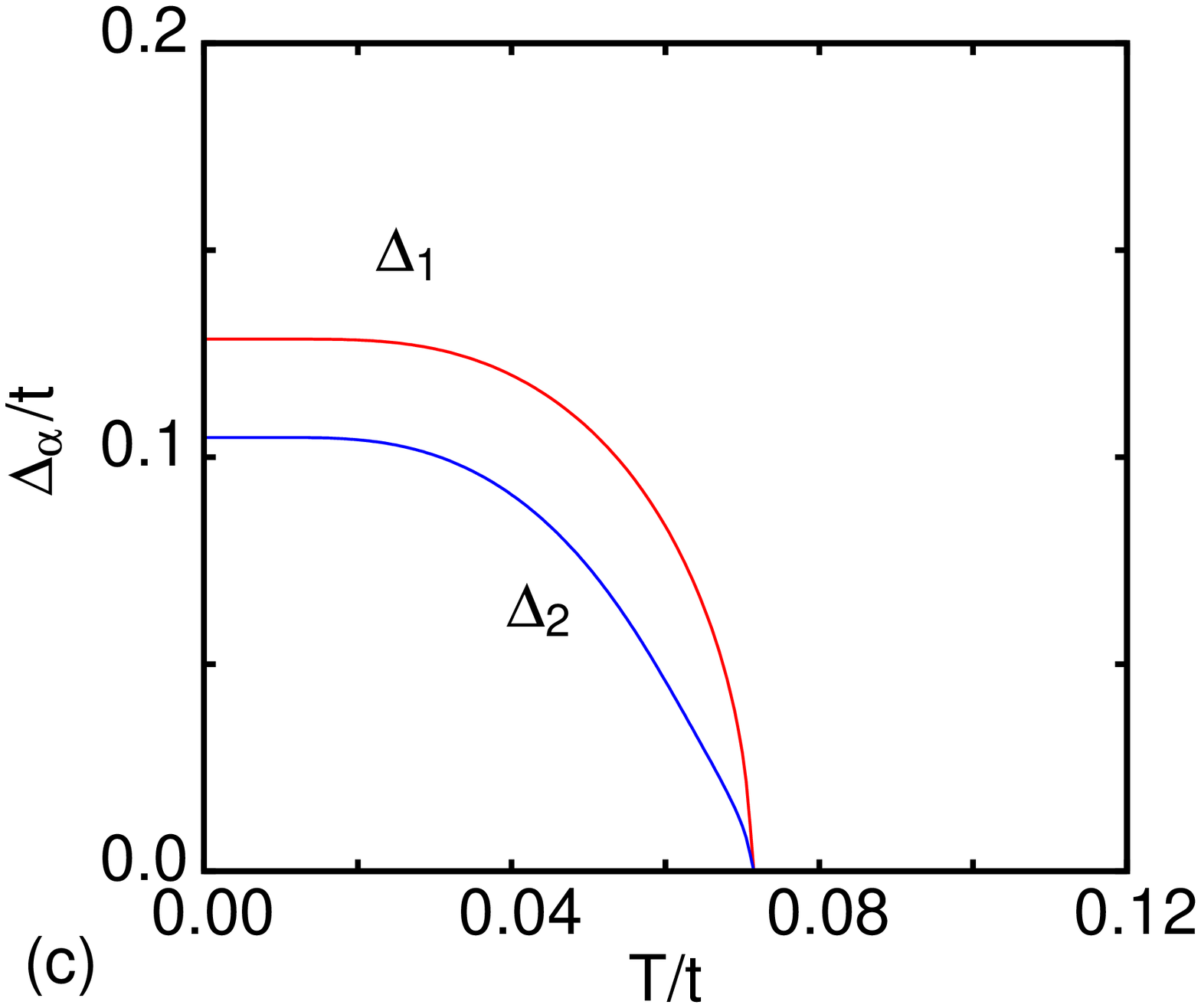} \hspace{-0.3cm}
\includegraphics[width=39mm]{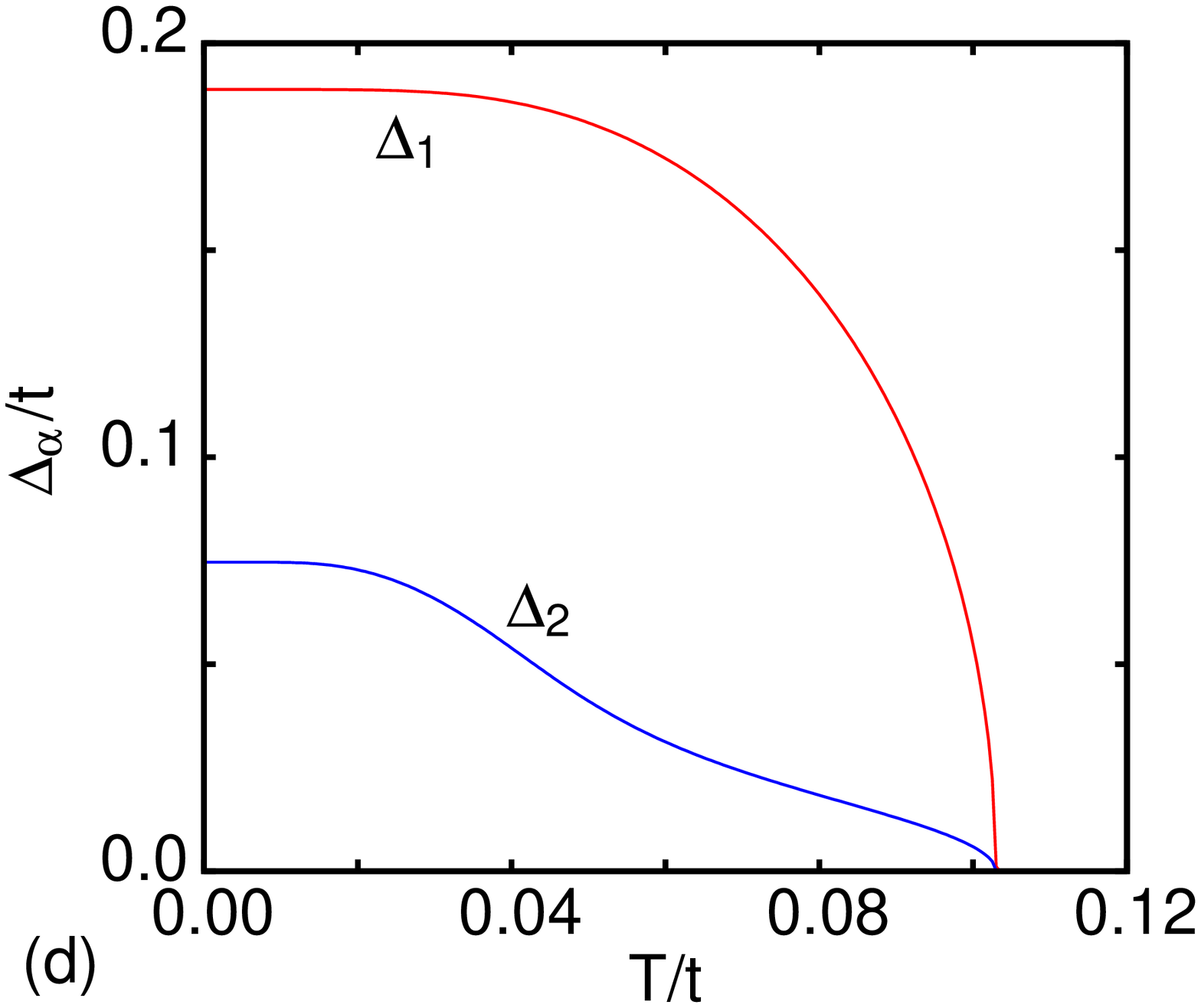}

\caption{
The temperature dependence of superconducting gaps for fixed $\varepsilon^{0}_1/t=2$ and $n= 1.10,
2.57, 2.64,
2.90$, panels (a)--(d), respectively.
}
%\end{center}
\end{figure}

Next we solve the gap equations (Eq. (\ref{3})) below $T_c$.
In Fig. 2 one can see the plotted superconducting gaps $\Delta_{\alpha}$ vs temperature for different band fillings. Note that the increase of  $n$ changes the order of $\Delta_1$ and $\Delta_2$ in the energy scale for fixed temperature, c.f. Figs. 2a, 2b and 2c, 2d. With that for the intermediate values on $n$ the gaps $\Delta_1$ and $\Delta_2$ are relatively close (Figs. 2b and 2c) while for $n=1.1$ and $n=2.9$ one of the gaps strongly dominates (Figs. 2a and 2d). The latter peculiarity is caused by the chemical potential
vicinity to the Van Hove singularity which supports superconductivity in the corresponding band. The weaker superconductivity in another band is related, at least partially, to the interband proximity effect. At the same time the band fillings $n=1.1$ and $n=2.9$ correspond approximately to the maxima of $T_{c}(n)$ in Fig. 1. Thus, the formation of two peaks of $T_{c}$ as a function of $n$ in Fig. 1 reflects the redistribution of the driving roles of bands in the appearance of superconductivity.

%fig3
\begin{figure}[!h]
%\begin{center}
\includegraphics[width=36mm,angle=-90]{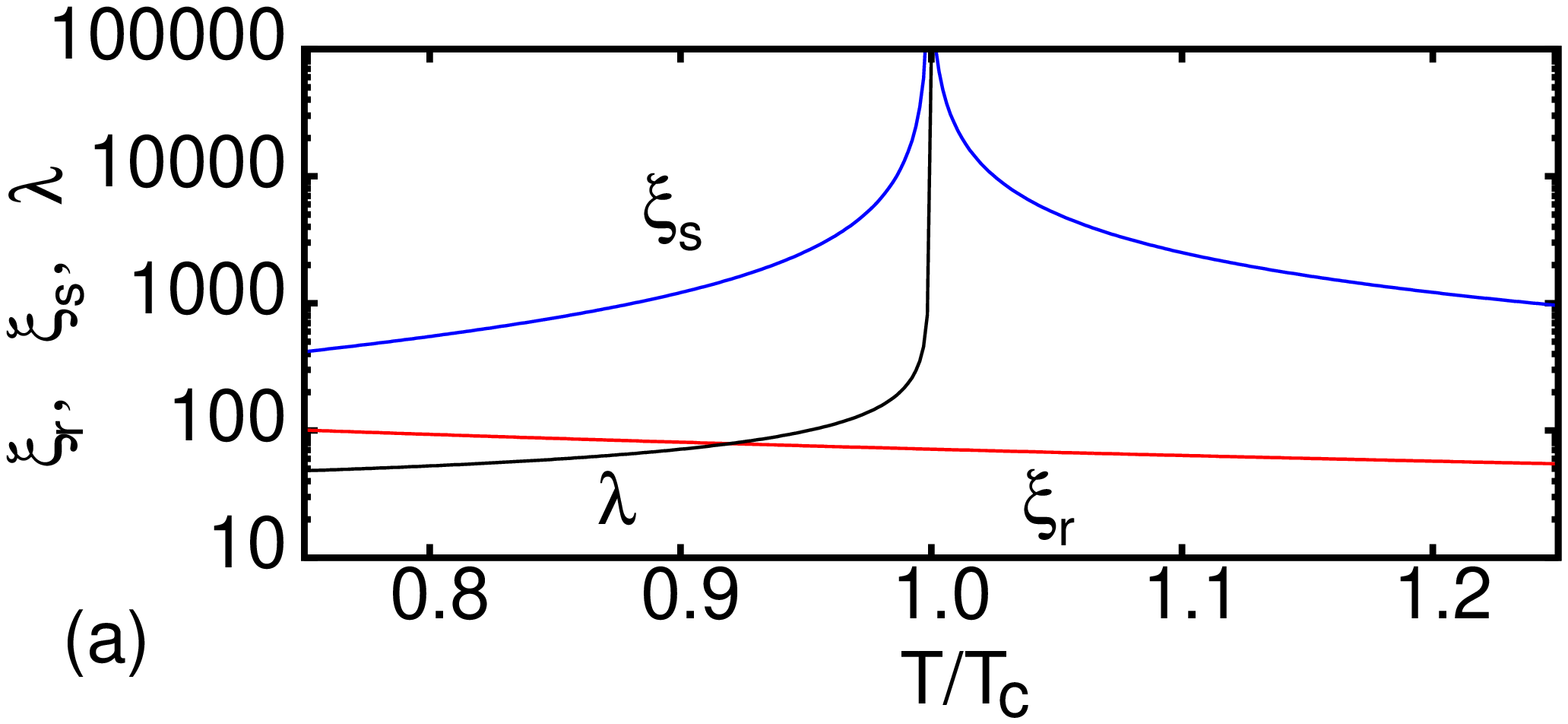}

\includegraphics[width=36mm,angle=-90]{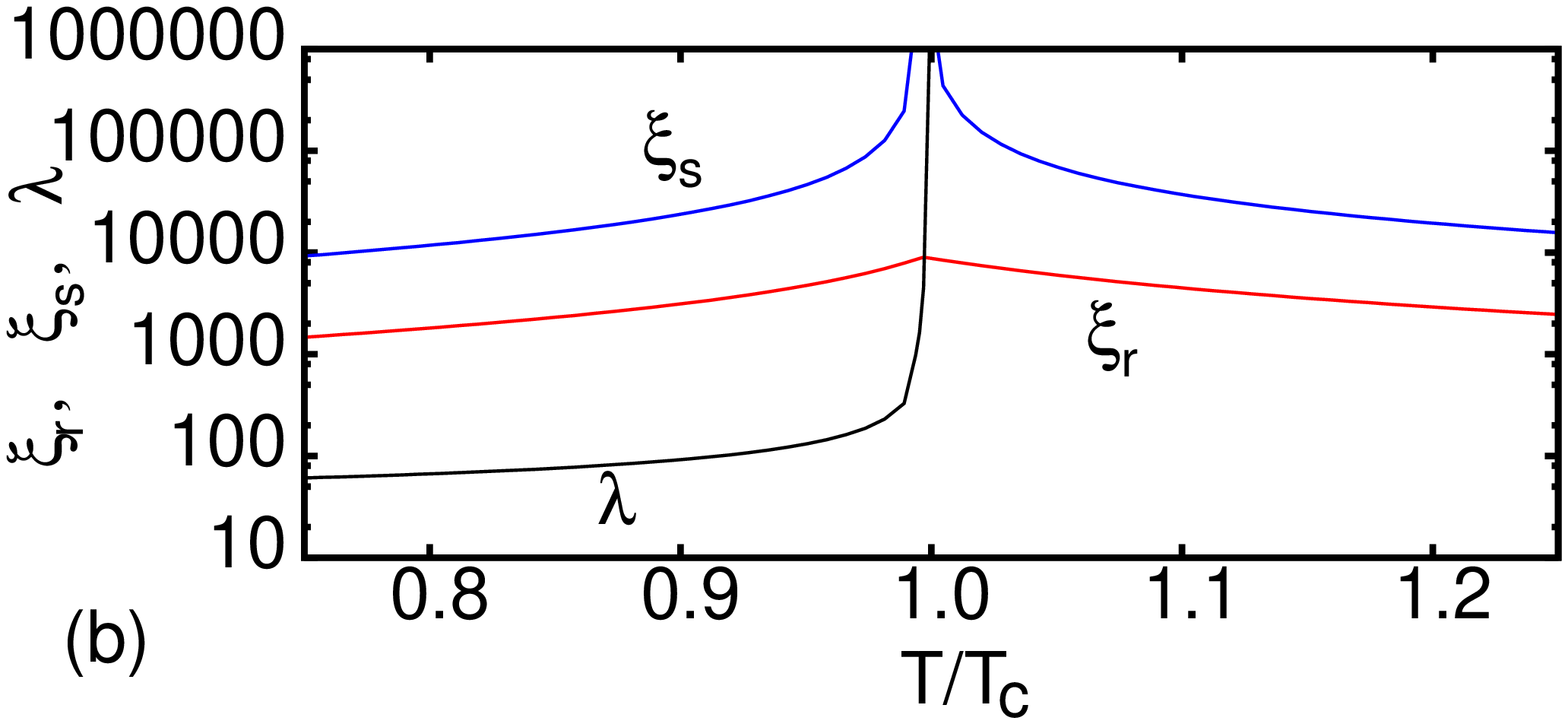}

\includegraphics[width=36mm,angle=-90]{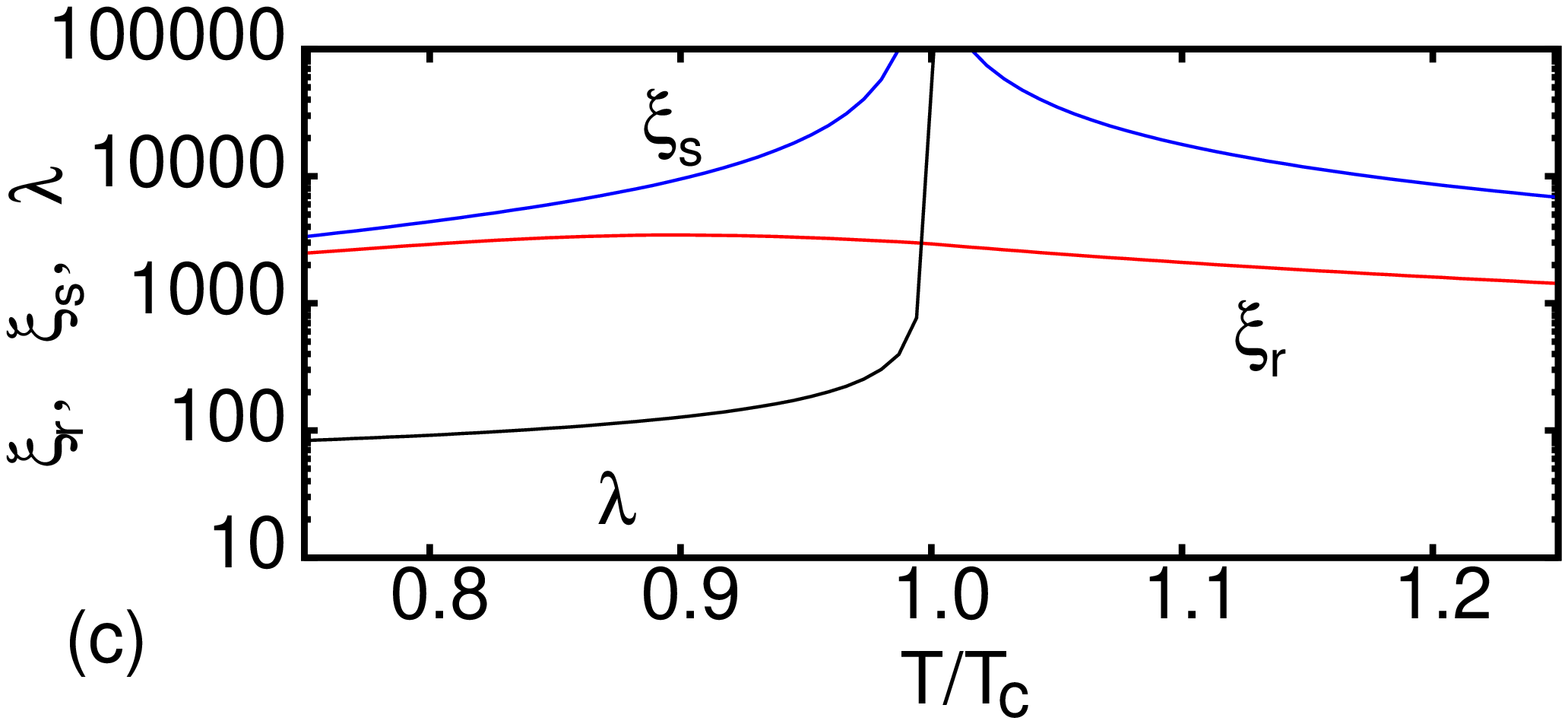}

\includegraphics[width=36mm,angle=-90]{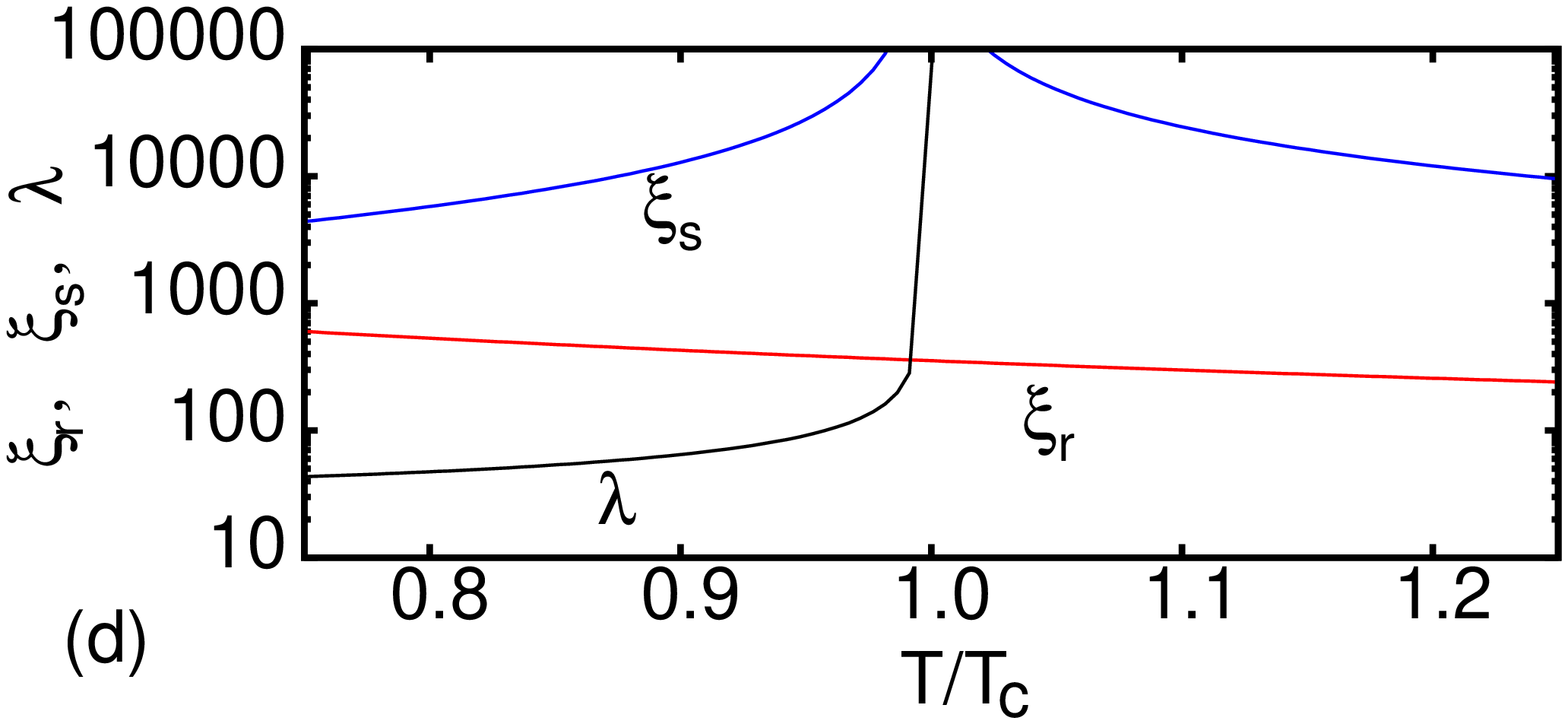}
\caption{
The temperature dependence of coherence lengths and magnetic field penetration depth in units of the
lattice space $a$
for fixed $\varepsilon^{0}_1/t=2$ and $n= 1.10, 2.57, 2.64, 2.90$, panels (a)--(d), respectively.}

%\end{center}
\end{figure}

In a single-orbital (single-band) system the Ginzburg-Landau parameter $\kappa=\lambda/\xi$ defines the type-I ($\kappa < 1/ \sqrt{2}$) and type-II ($\kappa > 1/ \sqrt{2}$) superconductivity. The coherence lengths (see Eq. (\ref{12})) and magnetic field penetration depth (see Eq. (\ref{16})) for two-orbital model are plotted in Fig. 3 for the band fillings discussed above. Apart from the divergence of the soft length scale $\xi_{s}$ at $T_{c}$, we observe also a maximum for the rigid length scale $\xi_{r}$ slightly below $T_{c}$ in Figs. 3b and 3c. This is due to the closeness of the superconductivity gaps for these band fillings, see Figs. 2b and 2c.

Another interesting observation is related to the crossing point of $\xi_r$ and $\lambda$. 
 In Figs. 3b-d this crossing appears at the temperatures very close to $T_c$ while in Fig. 3a the crossing point is located remarkably 
 lower in the temperature scale. Thus, there exists always a small temperature region in Figs 3a-d where $\xi_r < \lambda < \xi_s$ 
 \footnote{We have used the relatively small values of attractive interaction $|U^{\alpha\alpha}|$ to make the Hartree-Fock-Gorkov 
 approximation plausible.   In this situation we found that the penetration depth in our calculations is always smaller than the critical 
coherence length. The negative-U Hubbard model for larger $|U^{\alpha\alpha}|$ has to be treated in a different way  \cite{Ref mrr}.}.
One can treat the fulfilment of the inequality $\xi_r < \sqrt{2} \lambda
< \xi_s$ as necessary but not sufficient condition for type-1.5
superconductivity \cite{moschalkov}, see Ref. \cite{babaev}. To shed more
light on the problem one has to study the character of the interaction
between vortices \cite{babaev}. 

% One can treat the fulfillment of the condition $\xi_r < \sqrt{2} \lambda < \xi_s$ as the indication of type-1.5 superconductivity 
% \cite{moschalkov}, however, we stress here that only one length scale, $\xi_{s}$, can be attributed directly to the superconducting 
%phase 
% transition in a two-orbital model. To shed more light on the problem one has to study the vortex structure of that system carefully 
% \cite{babaev1}. 

The larger extent of the temperature region in Fig. 3a where $\xi_r < \lambda < \xi_s$ may be related to the interband 
 proximity effect which is more pronounced in this case (see Fig. 2a). It was shown in \cite{babaev2010} that type-1.5 superconductivity 
 can arise due to the proximity effect in a two-band system.

 %Finally, we note that the normalized superfluid density $\lambda^2(0)/\lambda^2(T)$ as a function of temperature varies very weakly for 
%the band fillings 
%considered, see Fig. 4. This scaling could be due to mean field approximation.

%fig4
%\begin{figure}[!h]
%\begin{center}
%\includegraphics[width=55mm,angle=-90]{sup_density.eps}
% \caption{Superfluid density, calculated for all the cases (a)-(d) denoted as in Figs. 2 and 3. Note that all the lines are very close.}
%\end{center}
%\end{figure}

\section{Conclusions}
We presented the results of band filling effect on two-orbital superconductor.
The dependencies of critical and non-critical coherence lengths on temperature were compared to the penetration depth for
different electron numbers. It was shown that the band structure including site energy and position of Van Hove singularity is important for superconducting gaps temperature behaviour as well as in the relationship between the magnitudes of critical and non-critical coherence lengths and penetration depth. We have identified a temperature region where $\xi_r < \lambda < \xi_s$. Its larger extent could be associated with the presence of interband proximity effect in a two-orbital superconducting system.

\section*{Acknowledgements}
 This research was supported by the European Union through the European Regional Development Fund (Centre of Excellence "Mesosystems: Theory and
Applications", TK114). We
acknowledge the support by the Estonian Science Foundation, Grant No 8991.
 GL kindly acknowledges a financial support by
the 7th Framework Programme FP7-REGPOT-2009-1,
under Grant Agreement No. 245479.

\end{document}